\documentclass[11pt,a4paper]{article}
\usepackage{epsfig}
\usepackage{graphics}

\newlength{\dinwidth}
\newlength{\dinmargin}
\def\eq#1{{Eq.~(\ref{#1})}}
\newcommand{\Le}{\left(}
\newcommand{\Ra}{\right)}

\newcommand{\beq}{\begin{equation}}
\newcommand{\eeq}{\end{equation}}
\newcommand{\beqar}{\begin{eqnarray}}
\newcommand{\eeqar}{\end{eqnarray}}
\newcommand{\D}{\partial}

\newcommand{\as}{\alpha_s}

\newcommand{\Ph}{\Phi}
\newcommand{\Php}{\Phi^{\dagger}}

\newcommand{\rad}{r^{2}_{12}}
\newcommand{\radd}{r^{2}_{23}}
\newcommand{\raddd}{r^{2}_{31}}

\newcommand{\nui}{i\nu}

\begin{document}

\title {{~}\\
{\Large \bf BFKL ansatz for BK equation in conformal basis }\\}
\,
\author{ {~}\\
{~}\\
{\bf S.Bondarenko\,\,${}^{a)}$\,
\thanks{Email: sergey@fpaxp1.usc.es},
\quad  A.Prygarin\,${}^{b),c)}$\,\thanks{E-mail: prygarin@post.tau.ac.il}}
\\[7mm]
{\it\footnotesize ${}^{a)}$ University of Santiago de Compostela,}\\
{\it\footnotesize Santiago de Compostela, Spain}\\
[7mm]
{\it\footnotesize ${}^{b)}$ HEP Department,  School of Physics and Astronomy,}\\
{\it\footnotesize Raymond and Beverly Sackler Faculty of Exact Science,}\\
{\it\footnotesize Tel-Aviv University, Ramat Aviv, 69978, Israel}\\
[7mm]
{\it\footnotesize ${}^{c)}$ II. Institut f\"ur Theoretische Physik, }\\
{\it\footnotesize Universit\"at Hamburg ,}\\
{\it\footnotesize Luruper Chaussee 149, D-22761~Hamburg, Germany}\\}
\maketitle
\thispagestyle{empty}

\begin{abstract}
The BK equation in the conformal basis is considered and analyzed. It is  shown that
at high energy a factorization of the coordinate and rapidity dependence should hold. This allows to simplify
significantly the from of the equation under discussion.
An analytical ansatz for the
solution to the BK equation at high energies is proposed and analyzed.
This analytical ansatz satisfies the initial condition at
low energy and does not depend on both rapidity and the initial condition in the high energy limit.
The case of the final rapidity being not too large is discussed and the properties of the transition region between small
and large final rapidities have been studied.

\end{abstract}

\section{Introduction}
The scattering of two distinct object, such as hadron and nucleus, or DIS at high energy is described by so-called "fan"
diagrams where only splitting of one pomeron into two  is taken into account \cite{vert1,vert2,brav}.
This situation was already discussed many years ago in the framework of the phenomenological pomeron \cite{Schwimmer}, and
some early attempts have been taken to generalize it for the case of QCD \cite{GLR}.
The equation for the  "fan" diagrams in the operator expansion formalism was written by I.Balitsky \cite{bal},
 and  in the dipole model framework \cite{muel} by Yu.Kovchegov \cite{kov}.
 The resulting Balitsky-Kovchegov (BK) equation is very well numerically studied in both, the saturation region and
 the region of small non-linearity (see  \cite{bksols,bdepkov,jimsol} and \cite{bksemi1,bksemi2}). One of the important
 features of the BK equation is the presence of the saturation scale at high energies that increases exponentially
 with rapidity, and a geometrical scaling of the solution \cite{barlev,scaling,traveling}. Despite the well
 understanding of the properties of the BK equation, the question of an exact analytical solution to the BK equation
 still remains open. This full  analytical solution may be very useful in the applications of the BK equation to
 different scattering processes at high energies.

In the present paper we want to bridge the gap in the analytical study of the BK equation. We consider the
formulation of the BK equation in the conformal basis and solve the problem using some simplifications.
In the conformal BK equation we keep only the terms with zero conformal spin, which corresponds to leaving only
 the BFKL pomerons propagators in the "fan" diagrams.
Due to highly complicated forms of the integrals appearing in the calculations the exact analytical solution is
beyond the scope of the present paper. However, we do discuss the possible ways of finding this exact analytical
 solution. We propose an ansatz of the solution to the BK equation, which relates between high energy behavior
 and the given initial condition at zero rapidity  for the pomeron field. This ansatz we call "phenomenological" or
 the BFKL ansatz for the solution to the BK equation. The reason for this is very simple, "phenomenological" because
 this solution is similar to phenomenological fan solution found in \cite{Schwimmer}
 (see also \cite{bon1} for more details) and the BFKL since the found solution corresponds to the leading BFKL
 terms in the "fan" structure leading at high energies.

The said ansatz was obtained using the conformal invariance of the pomeron field and the action of the
theory at high energy. The conformal invariance disappears when the final rapidity becomes not too large. This happens
due to the influence of the non-invariant source that cannot be neglected at low energy. Therefore, for small
final rapidities we consider a similar ansatz, which nevertheless does not possess
property of the conformal invariance. Having two solutions for two different regions we are able to analyze
the properties of the transition of the pomeron field from one region to another.

 The present paper is organized as follows. In the next section we consider the BK equation
 in the usual coordinate formulation and rewrite it in the conformal basis. In Sec. \ref{sec:scaling} we consider
 the conformal invariance of the theory at high energy and obtain scaling properties of the pomeron field.
In Sec. \ref{sec:solHighY} we propose a possible  ansatz  of solution to the BK equation in the conformal basis at high
energy and discuss the analytical structure of the solution. In Sec. \ref{sec:smallY} we consider an ansatz for the
pomeron field at small values of rapidity. Sec. \ref{sec:scales} is dedicated to the conditions and scales arising in the
region where the transformation of the pomeron field from the region of low energy to the region of the high energy
place.  In the last section  the conclusions and discussions are presented.

\section{BK equation in conformal basis}

 We begin our consideration with the effective field theory
of the interacting pomerons. It was shown by Braun ( see Refs.
\cite{braun1,braun2,braun3,braun4}) that one can describe the interacting pomerons in the large
$N_c$ limit  in terms of the effective action. The effective action with pomeron splitting only reads

\beq\label{EFT1}
S\,=\,S_0\,+\,S_I\,-\,S_E\,,
\eeq
where $S_0$ is a  free part of the action
\beq\label{EFT2}
\,S_0\,=\,\int\,dy\,dy^{'}\,d^{2}\rho_{1}\,d^{2}\rho_{2}\,d^{2}\rho^{'}_{1}\,
d^{2}\rho^{'}_{2}\,\Php(y,\rho_{1},\rho_{2})\,
G^{-1}_{y-y^{'}}(\rho_{1},\rho_{2}|\rho^{'}_{1},\rho^{'}_{2})\,
\Ph(y,\rho^{'}_{1},\rho^{'}_{2})\,
\eeq
$S_I$ is a interacting part of the action
\beq\label{EFT3}
S_I \,=\,\frac{2\,\as^2\,N_c}{\pi}\,
\int\,dy\,\int\,\frac{d^{2}\rho_{1}\,
d^{2}\rho_{2}\,d^{2}\rho_{3}}{\rad\,\radd\,\raddd}\,
\,(L_{13}\,\Php(y,\rho_{1},\rho_{3}))\,
\Ph(y,\rho_{1},\rho_{2})\,\Ph(y,\rho_{2},\rho_{3})\,
\,
\eeq
and $S_E$ is a source term of the action
\beq\label{AdEFT1}
S_E \,=\,\int\,dy\,\int\,d^{2}\rho_{1}\,
d^{2}\rho_{2}\,\Ph(y,\rho_{1},\rho_{2})\,\tau_{A}(y,\rho_{1},\rho_{2})\,+
\,\Php(y,\rho_{1},\rho_{2})\,\tau_{B}(y,\rho_{1},\rho_{2})\,
\eeq
The propagator of the theory
is defined through  the BFKL Hamiltonian \cite{lip1,Ham1,Ham2} as follows
\beq\label{EFT333}
G^{-1}_{y-y^{'}}(\rho_{1},\rho_{2}|\rho^{'}_{1},\rho^{'}_{2})\,=\,
\Le\,\nabla^{2}_{2}\,\nabla^{2}_{1}\,
\Le\,\frac{\D}{\D\,y}+H(\rho_{1},\rho_{2})\Ra\,\Ra\,
\delta^{2}(\rho_{1}\,-\,\rho^{'}_{1})\,
\delta^{2}(\rho_{2}\,-\,\rho^{'}_{2})\,
\delta(y-y^{'})\,\,
\eeq
The source terms in \eq{AdEFT1} are the initial forms of
the functions $\Ph$ and $\Php$ at rapidities $y=0$ and $y=Y$
respectively
\beq\label{AdEFT2}
\Ph(y,\rho_{1},\rho_{2})_{y=0}\,=\,
\tau_{B}(y,\rho_{1},\rho_{2})\,=\,\bar{\tau}_{B}(\rho_{1},\rho_{2})
\,\delta(y)\,
\eeq
\beq\label{AdEFT3}
\Php(y,\rho_{1},\rho_{2})_{y=Y}\,=\,
\tau_{A}(y,\rho_{1},\rho_{2})\,=\,\bar{\tau}_{A}(\rho_{1},\rho_{2})
\,\delta(y-Y)\,
\eeq
The pomeron field  $\Ph(y,r_{i},r_{j})$ in \eq{EFT2} and \eq{EFT3}
is the  generalized (skewed) gluon
(parton) distribution written in coordinate representation, see
\cite{bom,bon2},
and $L_{13}$ is the Casimir operator of the conformal group
\beq\label{EFT4}
L_{13}\,=\,\rho^{4}_{13}\,p^{2}_{1}\,p^{2}_{3} \,=\,\rho^{4}_{13}\,
\nabla^{2}_{1}\,\nabla^{2}_{3} \,\,
\eeq
In this formalism the BK equation is the equation of motion for
the pomeron field $\Ph(y,\rho_{i},\rho_{j})$
\beq\label{EFT5}
\Le  \frac{\D}{\D\,y}+H(\rho_{1},\rho_{3})\Ra
\,\Ph(y,\rho_{1},\rho_{3})\,+\,
\frac{2\,\as^2\,N_c}{\pi}\,
\int\,\frac{d^{2}\rho_{2}\,\rho^{2}_{31}}
{\rho^{2}_{12}\,\rho^{2}_{23}}\,
\,\Ph(y,\rho_{1},\rho_{2})\,\Ph(y,\rho_{2},\rho_{3})\,=\,0\,
\eeq
with the initial condition for the field given by \eq{AdEFT2}.

  For our further analysis it is more convenient to write \eq{EFT5}
in the conformal basis of functions $E_{\mu}(\rho_{1},\rho_{2})$ (see Ref.\cite{lip1})
\beq\label{EFT6}
E_{\mu}(\rho_{1},\rho_{2})\,=\,
\Le\,\frac{\rho_{12}}{\rho_{10}\,\rho_{20}}
\Ra\,^{h}\,
\Le\,\frac{\bar{\rho}_{12}}{\bar{\rho}_{10}\,\bar{\rho}_{20}}
\Ra\,^{\bar{h}}\,
\eeq
where $\mu=\{h,\rho_0\}$, $h=\frac{1+n}{2}+i\nu$ and $\bar{h}=1-h^{*}$.
For the sake of simplicity we adopted notation used in Ref.\cite{braun3}.
These functions are  the eigenfunctions of $L_{13}$ operator
\beq\label{EFT7}
L_{13}\,E_{\mu}(\rho_{1},\rho_{2})\,=\,
\lambda_{\mu}^{-1}\,E_{\mu}(\rho_{1},\rho_{2})\,
\eeq
with the eigenvalues
\beq\label{EFT8}
\lambda_{\mu}\,=
\frac{1}{[(n+1)^2\,+4\nu^2][ (n-1)^2\,+4\nu^2]}
\eeq
The pomeron field can be expanded in this basis as follows (see \cite{lip1})
\beq\label{EFT9}
\Ph(y,\rho_{1},\rho_{2})\,=\,\sum_{\mu}\,
E_{\mu}(\rho_{1},\rho_{2})\,
\Ph_ {\mu}(y)\,
\eeq
where
\beq\label{EFT10}
\Ph_ {\mu}(y)=
\int\,
\frac{d^{2}\,\rho_{1}\,d^{2}\rho_{2}}{\rho_{12}^{4}}\,
E_{\mu}^{*}(\rho_{1},\rho_{2})\,
\Ph(y,\rho_{1},\rho_{2})\,
\eeq
In \eq{EFT9} and hereafter by the conformal summation one should understand
\begin{eqnarray}\label{EFT11}
\sum_{\mu}\,=\,\sum_{n=-\infty}^{\infty}\,
\int_{-\infty}^{\infty}\,d\nu\,\frac{\nu^2\,+\,\frac{n^2}{4}}{\pi^4}\,
\int\,d^{2}\rho_{0}\,
\end{eqnarray}
Recasting  the Lagrangian of the theory in the conformal basis ( for more details see \cite{bon2}) one  obtains
\[
S\,=\,\int\,\,dy\,\sum_{\mu}\,
\left\{
\frac{1}{2}\,\Php_{\mu}(y)\,\lambda_{\mu}^{-1}\,
\frac{\D \Ph_{\mu}(y)}{\D y}\,-\,\frac{1}{2}\,
\Ph_{\mu}(y)\,\lambda_{\mu}^{-1}\,
\frac{\D \Php_{\mu}(y)}{\D y}\,-\,
\right.
\]
\beq\label{EFT12}
\left.
-\,\omega_{\mu}\lambda_{\mu}^{-1}
\Le\Ph_{\mu}(y)\,\Php_{\mu}(y)-\frac{2\as^2\,N_c}{\pi}
\omega_{\mu}^{-1} V_{\tilde{\mu},w,\nu}\,\Php_{\mu}(y)
\Ph_{w}(y)\Ph_{\nu}(y)\Ra
+\Ph_{\mu}(y)\,\tau_{\mu_{A}}\,+\,
\Php_{\mu}(y)\,\tau_{\mu_{B}}\,\right\}\,
\eeq
with  $\omega_{\mu}$  being  the BFKL eigenvalues
\beq\label{EFT122}
\omega_{\mu}=\omega_{h}=\bar{\alpha}\,
\Le\,\psi(1)-\mathrm{Re}\,
\psi(\frac{\mid\,n\,\mid\,+\,1}{2}\,+\,\nui)\,\Ra\,
\eeq
and withe the sources $\tau_{\mu_{A}}$ and $\tau_{\mu_{B}}$
for the fields $\Php$ and $\Ph$ at rapidities $y=Y$
and $y=0$ correspondingly.
The effective action in this form leads to
the equation of motion for each $\Ph_{\mu}$ field
\beq\label{EFT13}
\frac{\D \Ph_{\mu}(y)}{\D y}\,=\,
\omega_{\mu}\,\Ph_{\mu}(y)\,-\,\frac{2\as^2\,N_c}{\pi}\,\sum_{\mu_1,\mu_2}\,
\Ph_{\mu_1}(y)\Ph_{\mu_2}(y)\,
V_{\tilde{\mu},\mu_1,\mu_2}\,
\eeq
where $V_{\tilde{\mu},\mu_1,\mu_2}\,$
is the triple pomeron vertex in the conformal basis given by
\beq\label{EFT14}
V_{\mu,\mu_1,\mu_2}\,=
\int\,\frac{d^{2}\rho_{1}\,d^{2}\rho_{2}\,d^{2}\rho_{3}}
{\rho_{12}^{2}\,\rho_{23}^{2}\,\rho_{31}^{2}}\,
E_{\mu}(\rho_{1},\rho_{2})\,E_{\mu_1}(\rho_{2},\rho_{3})\,
E_{\mu_2}(\rho_{3},\rho_{1})\,\,
\eeq
and $\tilde{\mu}=\{1-h,\rho_0\}$.
The expression of \eq{EFT13} is, essentially,  the BK equation in the conformal
basis and
its solution is the main subject under consideration in the present paper.

Further simplifications of \eq{EFT14} using properties of the conformal group were performed in Ref.\cite{korm}
and the simplified expression reads
\beq\label{lead1}
V_{\mu_0,\mu_1,\mu_2}=
\Omega(h_{0},h_1,h_2)
(z_0-z_1)^{-\Delta_{01}} (z_1-z_2)^{-\Delta_{12}}
(z_2-z_3)^{-\Delta_{23}}
(\bar{z}_0-\bar{z}_1)^{-\bar{\Delta}_{01}} (\bar{z}_1-\bar{z}_2)^{-\bar{\Delta}_{12}}
(\bar{z}_2-\bar{z}_0)^{-\bar{\Delta}_{20}}
\eeq
where $\Delta_{12}=h_1+h_2-h_0$ and  $\bar{\Delta}_{12}=\bar{h}_1+\bar{h}_2-\bar{h}_0$.
We follow the notation of Ref.\cite{pesc,korm} and denote $\rho_0$ of $\mu_0$ by $z_0$, i.e.
 $\mu_0=\{h_0,z_0\}$ in \eq{lead1} and below.
Using the simplified expression for the triple pomeron vertex in \eq{lead1}
we can write
explicitly the nonlinear term in \eq{EFT13} as
\begin{eqnarray}\label{TripleA}
\frac{1}{4}\,\frac{2\as^2\,N_c}{\pi}\sum_{n_1,n_2=-\infty}^{\infty}\,\int_{-\infty}^{\infty}\,
d\nu_{1}\frac{\nu^{2}_{1}\,+\,\frac{n^{2}_{1}}{4}}{\pi^4}\,
\int_{-\infty}^{\infty}\,
d\nu_{2}\,\frac{\nu^{2}_{2}\,+\,\frac{n^{2}_{2}}{4}}{\pi^4}\,
\int\,d\,z_1\int\,d\,z_2\,
(z_0-z_1)^{-\Delta_{\tilde{0}1}}\,(z_1-z_2)^{-\Delta_{12}}\,
(z_2-z_0)^{-\Delta_{2\tilde{0}}}\,
\\
\times\int\,d\,\bar{z}_1\int\,d\,\bar{z}_2\,
(\bar{z}_0-\bar{z}_1)^{-\bar{\Delta}_{\tilde{0}1}}\,
(\bar{z}_1-\bar{z}_2)^{-\bar{\Delta}_{12}}\,
(\bar{z}_2-\bar{z}_0)^{-\bar{\Delta}_{2\tilde{0}}}\,\Ph_{\mu_1}(y,z_1)\,\Ph_{\mu_2}(y,z_2)\,\Omega(1-h_{0},h_1,h_2)\,
\nonumber
\end{eqnarray}

where the $\Delta_{\tilde{0}1}$    notation stands for $h_0$ to be replaced by $1-h_0$ in the expression for $\Delta_{01}$
defined above. This comes from the fact that the expression for \eq{EFT14} given by \eq{lead1} appears in \eq{EFT13} with
one of the conformal vertex functions conjugated.
To see the scaling properties of this nonlinear term we introduce  dimensionless variables
\begin{eqnarray}
w_1=\frac{z_1}{z_0}   \hspace{1cm},\hspace{1cm}   w_2=\frac{z_2}{z_0}
\end{eqnarray}
and their conjugate. In terms of these new dimensionless variables \eq{TripleA} reads

\begin{eqnarray} \label{Addlead3}
\frac{1}{4}\,\frac{2\as^2\,N_c}{\pi} \sum_{n_1,n_2=-\infty}^{\infty}\,\int_{-\infty}^{\infty}\,
d\nu_{1}\,\frac{\nu^{2}_{1}\,+\,\frac{n^{2}_{1}}{4}}{\pi^4}\,
\int_{-\infty}^{\infty}\,
d\nu_{2}\,\frac{\nu^{2}_{2}\,+\,\frac{n^{2}_{2}}{4}}{\pi^4}
\; z_{0}^{2-\Delta_{\tilde{0}1}-\Delta_{12}-\Delta_{2\tilde{0}}}\,
\,\bar{z}_{0}^{2-\bar{\Delta}_{\tilde{0}1}-\bar{\Delta}_{12}-
\bar{\Delta}_{2\tilde{0}}} \nonumber
\end{eqnarray}
\begin{eqnarray}
\times\int\,d\,w_1\int\,d\,w_2\,
(1-w_1)^{-\Delta_{\tilde{0}1}}\,(w_1-w_2)^{-\Delta_{12}}\,
(w_2-1)^{-\Delta_{2\tilde{0}}}\,
\int\,d\,\bar{w}_1\int\,d\,\bar{w}_2\,
(1-\bar{w}_1)^{-\bar{\Delta}_{\tilde{0}1}}\,
(\bar{w}_1-\bar{w}_2)^{-\bar{\Delta}_{12}}\,
(\bar{w}_2-1)^{-\bar{\Delta}_{2\tilde{0}}}\, \nonumber
\end{eqnarray}
\begin{eqnarray}
\times\Ph_{\mu_1}(y,z_{0}w_{1},\bar{z}_{0}\bar{w}_{1})\,
\Ph_{\mu_2}(y,z_{0}w_{2},\bar{z}_{0}\bar{w}_{2})\,\Omega(1-h_{0},h_1,h_2)\,
\end{eqnarray}
As one can see from \eq{Addlead3} the scaling properties of the triple pomeron term are merely determined form
the properties of the pomeron field and the power of the dimensionful  variable $z_0$.

\section{The scaling property of the pomeron field }\label{sec:scaling}
In this section we  show how the scaling properties of the pomeron field suggest
an ansatz for solution of the conformal BK equation at high energies.
 We come back to the scaling independent value of the
theory, to the  action given by \eq{EFT12}, and  check the scaling properties
of the pomeron field. We assume that our theory is conformal invariant
at high energies.
As a consequence of this fact the pomeron field scales accordingly to its dimensions
\beq\label{new77}
[\Ph_{\mu}(Y,z_{0})]=z^{h-1}_{0}\,\bar{z}^{\bar{h}-1}_{0}\,
\eeq
resulting in the following scaling property
\beq\label{new7}
\Ph_{\mu}(Y,\lambda\,z_{0},\bar{\lambda}\,\bar{z}_{0})\,=\,
\lambda^{h-1}\,\bar{\lambda}^{\bar{h}-1}\,\Ph_{\mu}(Y,z_{0},\bar{z}_{0})\,
\eeq
Next we analyze the scaling property of the source. We plug  the equation of motion
\eq{EFT13} into  the action of the theory and obtain
\beq\label{new1}
S\,=\,\int\,\,dy\,\sum_{\mu}\,\Ph_{\mu}(y)\,\tau_{\mu_{A}}(y)\,=\,
\sum_{n=-\infty}^{\infty}\,
\int_{-\infty}^{\infty}\,d\nu\,\frac{\nu^2\,+\,\frac{n^2}{4}}{\pi^4}\,
\int\,d^{2}\,z_{0}\,\,\Ph_{\mu}(Y,z_{0})\,\bar{\tau}_{\mu_{A}}(z_{0})\,,
\eeq
where Y is the final rapidity of the process and $\bar{\tau}_{\mu_{A}}(z_{0})$ is defined by
\beq\label{new2}
\tau_{\mu_{A}}(y)\,=\,\bar{\tau}_{\mu_{A}}(z_{0})\,\delta(y-Y)
\eeq

Using the scaling property of the pomeron field found in \eq{new7}
one can derive the scaling of the source
\beq\label{new8}
\lambda^{h}\,\bar{\lambda}^{\bar{h}}\,
\bar{\tau}_{\mu_{A}}(\lambda\,z_{0},\bar{\lambda}\,\bar{z}_{0})\,=\,
\bar{\tau}_{\mu_{A}}(z_{0},\bar{z}_{0})\,
\eeq
Another consequence of the scaling  property of the pomeron field
\eq{new7} is that it preserves unchanged the form of the equation of motion
with the triple pomeron term given by \eq{Addlead3}. This observation
and  the condition of the invariance of the action under the rescaling
prompts the following ansatz for the pomeron field at high energy
\beq\label{lead6}
\Ph_{\mu}(y,z)\,=\,z^{h-1}\,\bar{z}^{\bar{h}-1}\,
f^{(h,\bar{h})}(y)\,
\eeq
It is easy to see that  this ansatz satisfies all conditions  which we
have on the theory at high energy. Another important observation we made is the fact
that  \eq{lead6} presents a manifestation of the holomorphic separability of the pomeron field at high energies.

Now we are in position to  plug the ansatz \eq{lead6} into \eq{EFT13} with the help of \eq{TripleA} and read out the relevant
integrals over the coordinate variables. The coordinate integration factorizes and is performed in the Appendix A.
The resulting equation is the integro-differential equation for $f^{(h,\bar{h})}(y)$ function given by
\begin{eqnarray}\label{lead13}
\frac{\D f^{(h,\bar{h})}(y)}{\D y}&=&
\omega_{h}\,f^{(h,\bar{h})}(y)\,
-
\frac{\as^2\,N_c}{2\pi}\sum_{h_1,h_2}
f^{(h_1,\bar{h}_1)}(y)f^{(h_2,\bar{h}_2)}(y)\Omega(1-h,h_1,h_2)
I(1-h,h_1,h_2)I(1-\bar{h},\bar{h}_1,\bar{h}_2) \hspace{1cm}
\end{eqnarray}
where the integrals $I(1-h,h_1,h_2)$ and $I(1-\bar{h},\bar{h}_1,\bar{h}_2)$
are calculated in the Appendix A  and the summation over $h$ denotes
\begin{eqnarray} \label{sumH}
\sum_h=\sum_{n=-\infty}^{\infty}
\int d\nu
\frac{\nu^{2}+\frac{n^{2}}{4}}{\pi^4}
\end{eqnarray}
The  solution of \eq{lead13} would
give an exact analytical solution to the BK equation in the limit of high energy.

It is interesting to note that the form of the coordinate dependence of the function $\Phi_\mu(y)$  given by
the ansatz \eq{lead6}
resembles the form of the momentum dependence of the solution of
the BFKL equation for zero transferred momentum. The physical
meaning of this dependence is
different since the variables $z$ and $\bar{z}$ are conjugate to the transferred momentum while in the solution
of the BFKL equation such power dependence arises in the transverse momentum of the pomeron. However, the question
of a more subtle relation between the two cases is still to be  answered.
Another interesting observation, is that a factorized
form of the ansatz \eq{lead6} is a consequence of the fact that at high energy
the solution of the equation of motion does not "remember" about
the initial conditions and there is no another scale which can destroy
this factorization.

\section{Solution of BK equation in conformal basis at large final rapidities} \label{sec:solHighY}

 As it was already mentioned \eq{lead13} is our master one variable
integro-differential
equation, solution to which gives
the full analytical solution to the BK equation at high energy.
Unfortunately, being much simpler than the original BK equation
it still quite complicated. The exact solution of \eq{lead13} is beyond the scope of the present paper and we only
wish to find the leading solution at high energies. We consider the case of zero conformal spin $n_i=0$ which
corresponds to the fan diagrams with only the BFKL pomeron propagators. For this case \eq{lead13} reads
\begin{eqnarray}\label{sol1}
\frac{\D f^{\nu}(y)}{\D y}&=&
\omega_{\nu}\,f^{\nu}(y)\,
-
\frac{\as^2\,N_c}{2\pi}\int^{\infty}_{-\infty} \frac{\nu^2_1}{\pi^4}\;d\nu_1
\int^{\infty}_{-\infty} \frac{\nu^2_2}{\pi^4} \; d\nu_2
f^{\nu_1}(y)f^{\nu_2}(y)
\\&&\times\Omega\left(\frac{1}{2}-i\nu,\frac{1}{2}+i\nu_1,\frac{1}{2}+i\nu_2\right)
I^2(\frac{1}{2}-i\nu,\frac{1}{2}+i\nu_1,\frac{1}{2}+i\nu_2)
 \hspace{1cm}\nonumber
\end{eqnarray}
We want to investigate the properties of the solution to \eq{sol1} assuming that the function $f^{\nu}(y)$ can 
be expanded in powers of $\nu$
\beq\label{sol3}
f^{\nu}(y)=\sum_{k=0}^{\infty}\,f_{k}(y)\,\nu^{k}\,
\eeq
and plug into \eq{sol1}.  To the zeroth order we obtain 
\begin{eqnarray}\label{sol4}
\frac{\D f_0(y)}{\D y}&=&
\omega_{0}\,f_{0}(y)\,
-
\frac{\as^2\,N_c}{2\pi}f^2_{0}(y)\int^{\infty}_{-\infty} \frac{\nu^2_1}{\pi^4}\;d\nu_1
\int^{\infty}_{-\infty} \frac{\nu^2_2}{\pi^4} \; d\nu_2
\\&&\times\Omega\left(\frac{1}{2},\frac{1}{2}+i\nu_1,\frac{1}{2}+i\nu_2\right)
I(\frac{1}{2},\frac{1}{2}+i\nu_1,\frac{1}{2}+i\nu_2)I(\frac{1}{2},\frac{1}{2}+i\nu_1,\frac{1}{2}+i\nu_2)
 \hspace{1cm}\nonumber
\end{eqnarray}
or the same equation in more compact form
\beq\label{sol5}
\frac{\D f_{0}(y)}{\D y}\,=\,
\omega_{0}\,f_{0}(y)\,-
\frac{\as^2\,N_c}{2\pi}\,f_{0}^{2}(y)\,\mathbf{C}
\eeq
with some coefficient
\beq\label{sol6}
\mathbf{C}\,=\,
\int^{\infty}_{-\infty} \frac{\nu^2_1}{\pi^4}\;d\nu_1
\int^{\infty}_{-\infty} \frac{\nu^2_2}{\pi^4} \; d\nu_2
\; \Omega\left(\frac{1}{2},\frac{1}{2}+i\nu_1,\frac{1}{2}+i\nu_2\right)
I^2(\frac{1}{2},\frac{1}{2}+i\nu_1,\frac{1}{2}+i\nu_2)
\eeq
The solution to \eq{sol5} is readily found
\beq\label{sol7}
f_{0}(y)\,=\,\frac{e^{\omega_{0}\,y}}{Coeff+
\frac{\as^2\,N_c}{2\pi}\,
\frac{\mathbf{C}}{\omega_{0}}\,e^{\omega_{0}\,y}}\,
\eeq
where
the unknown coefficient $\,Coeff\,$ is to be determined from the initial
conditions  for $f_{0}(y)$ function.
Let's now simply write this condition in the following form
\beq\label{sol8}
f_{0}(y=0)\,=\,F_{in}\,
\eeq
which leads to
\beq\label{sol9}
f_{0}(y)\,=\,\frac{\,\exp(\omega_{0}\,y)\,F_{in}}
{\,\frac{\as^2\,N_c}{2\pi}\,F_{in}\,\frac{\mathbf{C}}{\omega_{0}}\,
\Le\,\exp(\omega_{0}\,y)-1\,\Ra\,+\,1\,}\,.
\eeq
We see, that at asymptotically high energy this ansatz does not depend
on the form of $\,F_{in}\,$, leaving the possibility for arbitrary
form of $\,F_{in}\,$. We will precisely define this function latter.
For obvious reasons we can call \eq{sol9} the double leading ansatz to
the solution of the BK equation. First, because
we consider only the leading term dominant at high energies, which is the BFKL fan structure; second, the function
$f_0(y)$ is the first term in the expansion in the powers of $\nu$ .
For the reasons we can call \eq{sol9} the  BFKL ansatz
for BK equation. This ansatz is also of a great interest since
it
is  similar to the solution to the analog of the BK equation
in the phenomenological pomeron theory, see for example \cite{bon1}.

  As a next step in our analysis we partially reconstruct $\nu$ dependence
of the function $f^{\nu}(y)$.  The form of \eq{sol9} suggests the following form of the solution with explicit
$\nu$ dependence
\beq\label{sol10}
f^{\nu}(y)\,=\,\frac{\,\exp(\omega_{\mu}\,y)\,F_{in}\,}
{\,F_{in}\,g_{0}^{-1}(\nu)\,
\Le\,\exp(\omega_{\mu}\,y)-1\,\Ra\,+\,1\,}\,,
\eeq
where $g_0(\nu)$ is the first term in the expansion of $f^{\nu}(y)$ in inverse powers of one pomeron exchange
\beq\label{sol12}
f^{\nu}(y)\,=\,
\sum_{k=0}^{\infty}\,g_{k}(\nu)\,e^{-k\,\omega_{\nu}\,y}\,.
\eeq
The function $g_{n}(\nu)$ are found from the equation \eq{sol1} in the form of
\begin{eqnarray}\label{sol101}
\omega_{\nu}(k+1)\,g_{k}(\nu)e^{-k\,\omega_{\nu}\,y}\,&=&
 \sum_{k_1,k_2=0}^{\infty}\frac{\as^2\,N_c}{2\pi}\int^{\infty}_{-\infty}
\frac{\nu^2_1}{\pi^4}\;d\nu_1
\int^{\infty}_{-\infty} \frac{\nu^2_2}{\pi^4} d\nu_2
\; g_{k_1}(\nu_1) e^{-k_1\,\omega_{\nu_1}\,y}\;  g_{k_2}(\nu_2)
e^{-k_2\,\omega_{\nu_2}\,y}\nonumber\\&&
\times\Omega\left(\frac{1}{2}-i\nu,\frac{1}{2}+i\nu_1,\frac{1}{2}+i\nu_2\right)
I^2(\frac{1}{2}-i\nu,\frac{1}{2}+i\nu_1,\frac{1}{2}+i\nu_2)
 \hspace{1cm}
\end{eqnarray}
As it was already mentioned due to complexity of $\nu$ dependent functions in
the kernel of this equation we are only able
 to find the  solution leading in energy. This means that we keep only terms with $k$,
$k_1$ and $k_2$ equal to zero
 obtaining equation for $g_0(\nu)$ only
\beq\label{sol11}
\omega_{\mu_{1}}\,g_{0}(\nu_{1})\,=\,
\frac{\as^2\,N_c}{2\pi}\,
\int_{-\infty}^{\infty}\,d\nu_{2}\,
\frac{\nu^{2}_{2}}{\pi^4}\,
\int_{-\infty}^{\infty}\,d\nu_{3}\,
\frac{\nu^{2}_{3}}{\pi^4}\,
g_{0}(\nu_{2})\,g_{0}(\nu_{3})\,\Omega\,I^2\,
\eeq
Plugging \eq{sol10} into \eq{lead6} we obtain, that when final rapidity of the
process is large, the asymptotically leading solution of BK equation
may be represented with the help of our ansatz
\beq\label{sol17}
\Ph_{\mu}(y)\,=|z_0|^{-1+2\,i\,\nu}\,\frac{\,\exp(\omega_{\mu}\,y)\,F_{in}\,}
{\,F_{in}\,g_{0}^{-1}(\nu)\,
\Le\,\exp(\omega_{\mu}\,y)-1\,\Ra\,+\,1\,}\,
\eeq
As it must be,  at the asymptotically large rapidity the pomeron
field \eq{sol17} can be written as
\beq\label{sol16}
\Ph_{\mu}(y)\,=|z_0|^{-1+2\,i\,\nu}\,g_{0}(\nu)\,,
\eeq
that confirmed the scaling property of the pomeron field at large
rapidities. In the case of the  rapidity smaller then the final rapidity
$Y$ of the process, the factorization
determined by \eq{lead6} is already not necessary and may be broken.
The possibility of the scenario when the final rapidity $Y$
is small is  discussed below.


 The ansatz \eq{sol10} correctly reproduce the property
of BK equation at large rapidities, namely the independence
of the ansatz on the initial conditions, the constant
behavior at high energy and scaling
independence of the action for the pomeron field.
Nevertheless, the function \eq{sol10} being used in the amplitude
only at rapidity $y=Y$, could be defined at all values of rapidity.
We find the form of $F_{in}$ function
from \eq{sol17} taking $y=0$ in the ansatz
\beq\label{rap1}
\Ph_{\mu}(y)_{y=0}\,=|z_0|^{-1+2\,i\,\nu}\,\,F_{in}\,
\eeq
As it was mentioned in the first section of the paper,
the Lagrangian of the theory includes the source terms. This implies
\beq\label{rap2}
\Ph_{\mu}(y)\,\delta(y)=\,\bar{\tau}_{\mu_{B}}\,\delta(y)
\eeq
resulting in
\beq\label{rap3}
|z_0|^{-1+2\,i\,\nu}\,\,F_{in}\,=\,\bar{\tau}_{\mu_{B}}\,
\eeq
Plugging $F_{in}$ from \eq{rap3} back into \eq{sol10}
we obtain for our ansatz
\beq\label{rap4}
\Ph_{\mu}(y)\,=\,\frac{\,\exp(\omega_{\mu}\,y)\,\bar{\tau}_{\mu_{B}}\,}
{\,|z_0|^{1-2\,i\,\nu}\,\bar{\tau}_{\mu_{B}}\,g_{0}^{-1}(\nu)\,
\Le\,\exp(\omega_{\mu}\,y)-1\,\Ra\,+\,1\,}\,
\eeq
The analytical ansatz of \eq{rap4}
interpolates between two desirable features of the behavior
of the solution to BK equation. At high rapidity it is determined by the
function which does not dependent on
rapidity and initial condition of the problem, whereas
at rapidity zero it is equal to the given initial function of the
pomeron field. It is important to underline therefore,
that the source field $\bar{\tau}_{\mu_{B}}$ in \eq{rap4} is arbitrary due these properties of the ansatz.
It is washed out at high rapidity
and there are no special constraints on the functional form of this source.
This source function may be arbitrary and may depend on some external scales of the problem.

\section{Solution to BK equation in conformal basis at small final rapidities}\label{sec:smallY}

In this section we consider a solution at small final rapidity $Y$.
In this case the asymptotic expansion in powers of $e^{-\omega_{\mu}\,y}$ in \eq{sol17} is not valid
anymore. The proposed ansatz given in \eq{rap4} does not describe correctly the solution to the BK equation and does not provide
the scaling invariant solution for the amplitude as well.
It is not surprising since we do not expect that
at low final rapidity the solution will preserve this
scaling invariance property.
Indeed, let us return to the triple pomeron
term in the equation of motion \eq{Addlead3}, keeping
only zero conformal spins in the formulae. Clearly,
all our previous consideration are  valid if we can
justify the asymptotic expansion of the pomeron field in the form
\beq\label{sm1}
\Ph_{\mu}(y)\,=\,\sum_{k=0}^{\infty}\,g_{k}^{\mu}(z)\,
e^{-k\omega_{\mu}\,y}\,
\eeq
However, if $Y$ is small   another expansion should hold
\beq\label{sm2}
\Ph_{\mu}(y)\,=\,\sum_{k=1}^{\infty}\,\bar{g}_{k}^{\mu}(z)\,
e^{k\omega_{\mu}\,y}\,.
\eeq
Similar the previous case we  put the expansion of  \eq{sm2} into the
equation of motion keeping only the first term of this expansion. It is easy to see
that at small rapidities the triple pomeron term is not enhanced
by rapidity exponential and can be safely neglected. In this case we obtain the BFKL pomeron solution
which corresponds to the first term in the expansion \eq{sm2}
\beq\label{sm3}
\Ph_{\mu}(y)\,=\,\bar{\tau}_{\mu_{B}}\,e^{\omega_{\mu}\,y}\,
\eeq
and  determines the form of the action
\beq\label{sm4}
S\,=\,
\int_{-\infty}^{\infty}\,d\nu\,\frac{\nu^2\,}{\pi^4}\,
\int\,d^{2}\,z_{0}\,\bar{\tau}_{\mu_{B}}(z_{0})\,e^{\omega_{\mu}\,y}\,
\bar{\tau}_{\mu_{A}}(z_{0})\,
\eeq
Similar to the case of the pomeron field we deduce from the from of the action the scaling property
of the
$\bar{\tau}_{\mu_{B}}$, namely,
\beq\label{sm7}
\bar{\tau}_{\mu_{B}}(\lambda\,z_{0},\bar{\lambda}\,\bar{z}_{0})\,=\,
\lambda^{-1+2\,i\nu}\,
\bar{\tau}_{\mu_{B}}(z_{0},\bar{z}_{0})\,
\eeq
 Naturally, for an   arbitrary source field $\bar{\tau}_{\mu_{B}}$
such a strong constraint, in general,  is not  satisfied.
Therefore, we do not expect the invariance of the
action under the rescaling  of the variable $z_0$ at small  energies.
Such an invariance restores at large energies, when the dependence
of the amplitude on the source $\bar{\tau}_{\mu_{B}}$  disappears.

 As a next step in our discussion we  consider the first  two  terms in  the expansion \eq{sm2}
of the pomeron field at small energy
\beq\label{nsm1}
\Ph_{\mu}(y)\,=\,(\,\bar{\tau}_{\mu}(z_{0})\,-\,\bar{g}_{1}^{\mu}(z_{0})\,)
e^{\omega_{\mu}\,y}\,+\bar{g}_{1}^{\mu}(z_{0})\,e^{2\,\omega_{\mu}\,y}\,.
\eeq
Using equation of motion we obtain the equation for
the $\bar{g}_{1}^{\mu}(z_{0})$ function from this expansion
\beq\label{nsm2}
\bar{g}_{1}^{\mu}(z_{0})\,e^{2\omega_{\mu}\,y}\,\omega_{\mu}
=\,-\,
\frac{2\as^2\,N_c}{\pi}\,\sum_{\mu_1,\mu_2}\,
V_{\tilde{\mu},\mu_1,\mu_2}\,
(\bar{\tau}_{\mu_{1B}}(z_{1})\,-\,\bar{g}_{1}^{\mu_{1}}(z_{1})\,)\,
(\bar{\tau}_{\mu_{2B}}(z_{2})\,-\,\bar{g}_{1}^{\mu_{2}}(z_{2})\,)
e^{(\omega_{\mu_1}\,+\,\omega_{\mu_2})y}\,
\eeq

In this way we can consider an expansion with any arbitrary number of terms obtaining a chain of the
equations similar to \eq{nsm2} with involved interference terms between the functions $\bar{g}_{i}^{\mu}(z_{0})$.
Instead, we find an ansatz that have a structure of the expansion \eq{sm2} and will coincide with the expansion
\eq{sm2} to some order. As the simplest example we take expansion up to the second order and the form of this
ansatz is borrowed from \eq{rap4} as follows
\beq\label{sm8}
\Ph_{\mu}(y)\,=\,\frac{\,\exp(\omega_{\mu}\,y)\,\bar{\tau}_{\mu_{B}}\,}
{\,|z_0|^{1-2\,i\,\nu}\,\bar{\tau}_{\mu_{B}}\,\bar{g}_{0}(\nu,z_0)\,
\Le\,\exp(\omega_{\mu}\,y)-1\,\Ra\,+\,1\,}\,
\eeq
Expanding \eq{sm8} and comparing the first two
terms of this expansion to  the \eq{nsm1} we obtain
\beq\label{nsm3}
\bar{g}_{1}^{\mu}(z_{0})\,=\,-\,
|z_0|^{1-2\,i\,\nu}\,\bar{\tau}_{\mu_{B}}^{2}\,\bar{g}_{0}(\nu,z_0)\,
\eeq
We use this expression in \eq{nsm2} to obtain the equation for the function $\bar{g}_{0}(\nu,z_0)\,$, namely,
\[
|z_0|^{1-2\,i\,\nu}\,\bar{\tau}_{\mu_{B}}^{2}\,\bar{g}_{0}(\nu,z_0)\,
\,e^{2\omega_{\mu}\,y}\,\omega_{\mu}
=\,
\]
\beq\label{nsm4}
\frac{2\as^2\,N_c}{\pi}\,\sum_{\mu_1,\mu_2}
V_{\tilde{\mu},\mu_1,\mu_2}
\bar{\tau}_{\mu_{1B}}(z_{1})\bar{\tau}_{\mu_{2B}}(z_{2})
(1-|z_1|^{1-2\,i\,\nu_1}\,\bar{\tau}_{\mu_{1B}}\,\bar{g}_{0}(\nu_1,z_1))
(1-|z_2|^{1-2\,i\,\nu_2}\,\bar{\tau}_{\mu_{2B}}\,\bar{g}_{0}(\nu_2,z_2))
e^{(\omega_{\mu_1}+\omega_{\mu_2})y}\,.
\eeq
Further simplification of the obtained equations is possible if one assumes that
\beq\label{nsm5}
1\,<\,|z|^{1-2\,i\,\nu}\,\bar{\tau}_{\mu_{B}}\,\bar{g}_{0}(\nu,z)\,.
\eeq
in the expansion \eq{sm2}.
This condition means a smallness of the source multiplied by the triple
pomeron vertex function. Indeed, there is a suppression of this term
compared  to unity
due to $\alpha_s^2\,$ in front of \eq{nsm4} if the source
does not contain some large number (number of nucleons in nucleus, for example).
In this case expanding the pomeron field \eq{sm8}
one obtains
\beq\label{nsm6}
\Ph_{\mu}(y)\,=\,\exp(\omega_{\mu}\,y)\,\bar{\tau}_{\mu_{B}}\,
-\,|z_0|^{1-2\,i\,\nu}\,\bar{\tau}_{\mu_{B}}^{2}\,\bar{g}_{0}(\nu,z_0,y)\,
\,e^{2\omega_{\mu}\,y}\,
\eeq
and  the equation \eq{nsm4} becomes
\beq\label{nsm7}
|z_0|^{1-2\,i\,\nu}\,\bar{\tau}_{\mu_{B}}^{2}\,\bar{g}_{0}(\nu,z_0,y)\,
\,e^{2\omega_{\mu}\,y}\,\omega_{\mu}
=
\frac{2\as^2\,N_c}{\pi}\,\sum_{\mu_1,\mu_2}
V_{\tilde{\mu},\mu_1,\mu_2}\,
\bar{\tau}_{\mu_{1B}}(z_{1})\,\bar{\tau}_{\mu_{2B}}(z_{2})\,
e^{(\omega_{\mu_1}+\omega_{\mu_2})y}\,
\eeq
In \eq{nsm7} we  introduced a rapidity dependence
in the function $\bar{g}_{0}$
as the price for this simplification
\beq\label{nsm8}
\bar{g}_{0}(\nu,z_0)\,\rightarrow\,\bar{g}_{0}(\nu,z_0,y)\,
\eeq
since in \eq{nsm7} this function possess  some
subdominant rapidity corrections.
Thus in the case when \eq{nsm7} may be used instead \eq{nsm4}
our ansatz becomes
\beq\label{nsm9}
\Ph_{\mu}(y)\,=\,\frac{\,\exp(\omega_{\mu}\,y)\,\bar{\tau}_{\mu_{B}}\,}
{\,|z_0|^{1-2\,i\,\nu}\,\bar{\tau}_{\mu_{B}}\,\bar{g}_{0}(\nu,z_0,y)\,
\Le\,\exp(\omega_{\mu}\,y)-1\,\Ra\,+\,1\,}\,.
\eeq
The obtained equations are more complicated then that of \eq{sol17}
in the high rapidity limit,
because there is no factorization of $z$ and $\nu$
variables.

\section{The pomeron field at all rapidities and a transition region between small and large values of rapidity}\label{sec:scales}
The obtained ansatzs for large and small values of final rapidity indicate, that
the only function that will be changed due to the rapidity evolution
will be a $g$ function in the denominators of the ansatz.
Therefore, as a definition of the large and small rapidities we could use
a following observations.
Between these two rapidity regions of the solution, at asymptotically
large rapidity and small rapidity,
exists a region of the transition between growing and saturated
behavior of the pomeron field. In general, this region may be defined
as a region where we can instead
expansion in exponents $e^{\omega_{\mu}\,y}$ we use
a asymptotic expansion in the exponents $e^{-\omega_{\mu}\,y}$ and vice versa. Therefore, from the forms of
\eq{rap4} and \eq{sm8} in this region may be defined
by the following conditions
\beq\label{rap5}
\,|z_{cr}|^{1-2\,i\,\nu}\,\bar{\tau}_{\mu_{B}}(z_{cr})\,g_{0}^{-1}(\nu)\,
\Le\,\exp(\omega_{\mu}\,y_{cr})-1\,\Ra\,\propto\,1\,,
\eeq
and
\beq\label{rap55}
\,|z_{cr}|^{1-2\,i\,\nu}\,\bar{\tau}_{\mu_{B}}(z_{cr})\,\bar{g}_{0}(\nu,z_{cr})\,
\Le\,\exp(\omega_{\mu}\,y_{cr})-1\,\Ra\,\propto\,1\,
\eeq
This, in analogy with the usual definition of saturation momenta,
will define the critical scale of the $z_0$ variable in the conformal basis
\beq\label{rap6}
\,|z_{cr}|^{1-2\,i\,\nu}\,\bar{\tau}_{\mu_{B}}(z_{cr})\,
\propto\,g_{0}(\nu)\,
\Le\,\exp(\omega_{\mu}\,y_{cr})-1\,\Ra^{-1}\,.
\eeq
for which the transition occurs.
It is interesting to note, that the source
$\bar{\tau}_{\mu_{B}}(z_{cr})$ in \eq{rap6}  brings some
external scales dependence in \eq{rap6}, which  defines
the character  of transition.
We also could consider the constraint \eq{rap6} as a definition
of some "transition" rapidity $y_{cr}$ for fixed values of the final
rapidity $Y$, vector $z_{0}$ and
arbitrary external scales inside the source $\bar{\tau}_{\mu_{B}}$
\beq\label{rap7}
y_{cr}\,\propto\,\frac{1}{\omega_{\mu}}\,\ln\Le\,
1\,+\,g_{0}(\nu)\,|z_{cr}|^{-1+2\,i\,\nu}\,
\bar{\tau}_{\mu_{B}}^{-1}(z_{cr})\,\Ra\,.
\eeq
Comparing \eq{rap5} and \eq{rap55} we see
that one can define  the transition region as  a region, where
the  following relation is  satisfied
\beq\label{rap8}
\,g_{0}^{-1}(\nu)\,\approx\,\bar{g}_{0}(\nu,\,z_0)_{z_{0}=z_{cr}}\,
\eeq
This relation  shows, inter alia,  that if we find the
functional form of $\,\bar{g}_{0}(\nu,\,z_0)\,$, then
this function must satisfy the convergent expansion
in $z_0$ around some value $z_{cr}$ determined by \eq{rap8}
\beq\label{rap9}
\bar{g}_{0}(\nu,\,z_{0})\,=\,\sum_{n=0}^{\infty}\,g_{n}^{-1}(\nu)\,
(z_{0}-z_{cr})^{n}\,.
\eeq
So, with the use of the $\,\bar{g}_{0}(\nu,\,z_0)\,$
function we could find  the $\,g_{0}^{-1}(\nu)\,$
function through \eq{rap9} as well.
Of course, for that we need to know a form of the $\,\bar{g}_{0}(\nu,\,z_0)\,$
function, that for general case is not easy.
Therefore, instead, we could define as a transition region
the region where both terms of expansion \eq{nsm6} are equal
\beq\label{rap10}
\,\exp(\omega_{\mu}\,y)\,\bar{\tau}_{\mu_{B}}\,
=\,|z_0|^{1-2\,i\,\nu}\,\bar{\tau}_{\mu_{B}}^{2}\,\bar{g}_{0}(\nu,z_0,y)\,
\,e^{2\omega_{\mu}\,y}\,.
\eeq
Using the \eq{nsm7} we finally will obtain condition for the region where
two solutions are overlapping
\beq\label{rap11}
\,e^{\omega_{\mu}\,y_{cr}}\,\bar{\tau}_{\mu_{B}}(z_{cr})\,
=\,\frac{2\as^2\,N_c}{\pi}\,\sum_{\mu_1,\mu_2}
V_{\tilde{\mu},\mu_1,\mu_2}\,
\bar{\tau}_{\mu_{1B}}(z_{1})\,\bar{\tau}_{\mu_{2B}}(z_{2})\,
e^{(\omega_{\mu_1}+\omega_{\mu_2})y_{cr}}\,.
\eeq
Clearly, this is a screening condition on the sources
of the problem, which defines values of $z_{cr}$ and $y_{cr}$
for which the source of the projectile will be screened from the target
by the triple pomeron interactions.

\section{The accuracy of ansatz }\label{sec:pres}

The source of possible corrections to the pomeron field at
large rapidities are the coefficients $\,g_{k}(\nu)\,$
for different $k$ in the series expression for the pomeron field
\beq\label{pres2}
\Ph_{\mu}(y,z)\,=\,z^{h-1}\,\bar{z}^{\bar{h}-1}\,
\sum_{k=0}^{\infty}\,\,g_{k}(\nu)\,e^{-k\,\omega_{\nu}\,y}\,
\eeq

In the similar expansion for the phenomenological pomeron the ration of the coefficients of the successive terms
is proportional to $\alpha_s$. Based on this information we make an assumption that this is also
the case in the expansion \eq{pres2} and we only need to find the overall normalization, i.e. the order in $\alpha_s$
of the first term in the expansion. 
 
 We plug this series into the equation of motion and write the first two equation
from the chain of equations \eq{sol101}. In integrals over $\nu$ we assume a contribution
from such regions of $\nu$ in which  the BFKL structure of the series \eq{pres2} is kept,
namely, from the "diffusion regions" of $\nu$ with $e^{\omega_{\nu}\,y}\,\approx\,e^{\omega_{0}\,y}$.
In the first order of expansion in $e^{-k\,\omega_{\nu}\,y}$
we obtain \eq{sol11}
\beq\label{pres3}
\omega_{\mu_{1}}\,g_{0}(\nu_{1})\,=\,
\frac{\as^2\,N_c}{2\pi}\,
\int_{-\infty}^{\infty}\,d\nu_{2}\,
\frac{\nu^{2}_{2}}{\pi^4}\,
\int_{-\infty}^{\infty}\,d\nu_{3}\,
\frac{\nu^{2}_{3}}{\pi^4}\,
g_{0}(\nu_{2})\,g_{0}(\nu_{3})\,\Omega\,I^2\,
\eeq
One can easily see from counting the powers of $\alpha_s$ that  the coefficient function
$g_{0}(\nu)$ is of the order of $1/\alpha_s$.

To show what are the limitations for the accuracy of the proposed ansatz 
we expand the expression in \eq{rap4} for the large values of rapidity 
\beq\label{pres8}
\Ph_{\mu}(y,z)=z^{h-1}\bar{z}^{\bar{h}-1}\,g_{0}(\nu)\Le
1+e^{-\omega_{\mu}\,y}\Le\,1-\frac{z^{h-1}\bar{z}^{\bar{h}-1}}
{\bar{\tau}_{\mu}}\,g_{0}(\nu)\Ra\,+e^{-2\omega_{\mu}\,y}\Le\,1-\frac{z^{h-1}\bar{z}^{\bar{h}-1}}
{\bar{\tau}_{\mu}}\,g_{0}(\nu)\Ra^{2}\,+...\Ra
\eeq
or, approximately,
\beq\label{pres9}
\Ph_{\mu}(y,z)=z^{h-1}\bar{z}^{\bar{h}-1}\,g_{0}(\nu)\Le
1-e^{-\omega_{\mu}\,y}\,\frac{z^{h-1}\bar{z}^{\bar{h}-1}}
{\bar{\tau}_{\mu}}\,g_{0}(\nu)\,+\,e^{-2\omega_{\mu}\,y}\,
\Le\,\frac{z^{h-1}\bar{z}^{\bar{h}-1}}
{\bar{\tau}_{\mu}}\,g_{0}(\nu)\Ra^{2}\,+...\Ra
\eeq
provided we neglect subleading terms in $\alpha_s$ assuming  $\,1\,<\,1\,/\,\alpha_s\,$.
We see that the expansion  \eq{pres9} properly reproduces the expected 
$\,(\alpha_{s}^{k}\,e^{(k-1)\omega\,y})^{-1}\,$ behavior for each term, in the agreement 
with that of the phenomenological pomeron.

Using this $\alpha_s$ structure of the expansion \eq{pres9} one can easily see 
that the series converges for rapidities in the region of 
\beq\label{pres5}
y\,>\,\frac{1}{\alpha_s}\,\ln(\frac{1}{\alpha_s})\,
\eeq
Based on our discussion we conclude that the proposed ansatz \eq{nsm9} has both the correct
high energy behavior and the leading $\alpha_s$ expansion in the rapidity region given by \eq{pres5}. 
If one wishes to take into account higher order corrections in $\alpha_s$ in the expansion 
\eq{pres2}, the proper way to do this is to introduce a new form of the ansatz

\beq\label{pres11}
\Ph_{\mu}(y)\,=\,\frac{\,\exp(\omega_{\mu}\,y)\,\bar{\tau}_{\mu_{B}}\,}
{\,|z_0|^{1-2\,i\,\nu}\,\bar{\tau}_{\mu_{B}}\,F\,
\Le\,\exp(\omega_{\mu}\,y)-1\,\Ra\,+\,1\,}\,
\eeq
with some function
\beq\label{pres12}
F\,=\,\sum_{n=-1}^{\infty}\,\alpha_{s}^{n}\,f_{n}(\nu,y)
\eeq
where the all higher corrections are encoded.

It is a straightforward procedure to show in a similar way that the
 same arguments
also hold for the low rapidity expansion where one expands the ansatz in powers of $e^{\omega_{\mu}y}$.
In this case the validity of the expansion is restricted to the rapidity region given by 
\beq\label{pres13}
y\,<\,\frac{1}{\alpha_s}\,\ln(\frac{1}{\alpha_s})\,
\eeq
 In order to  account for higher  order corrections in $\,\alpha_{s}\,$ 
in the ansatz \eq{sm8} one can also make a use of some function $F\,$
in a way similar to that  of  \eq{pres11}. However, in this case the function $F\,$  
possesses a different  from \eq{pres12} form of expansion in the powers of 
$\,\alpha_{s}\,$
\beq\label{pres14}
F\,=\,\sum_{n=1}^{\infty}\,\alpha_{s}^{n}\,f_{n}(\nu,y)
\eeq

\section{Conclusion}
In the present paper we discussed possible ways of an analytical solution to the BK equation
in the conformal basis. We suggested the following ansatz for the solution of BK equation

\beq\label{conc1}
\Ph_{\mu}(y)\,=\,\frac{\,\exp(\omega_{\mu}\,y)\,\bar{\tau}_{\mu_{B}}\,}
{\,|z_0|^{1-2\,i\,\nu}\,\bar{\tau}_{\mu_{B}}\,F\,
\Le\,\exp(\omega_{\mu}\,y)-1\,\Ra\,+\,1\,}\,
\eeq
where the form of the function $F$ in the denominator of \eq{conc1}
depends on the region of rapidity where the solution is considered.
The problem, therefore, is reduced to the evolution of this
unknown  function $F$ with rapidity.
Assuming the conformal invariance of the theory at high energy we simplify our problem proposing factorization of
the coordinate dependence of the pomeron field in the this limit. This makes it possible to separate the coordinate
and rapidity dependence resulting into \eq{lead13}. This equation is still not easy to solve, but it allows us to
investigate the energy dependence of the solution to the BK equation not mixing it with the coordinate degrees
 of freedom.
  The next important simplification we make is keeping only the BFKL structure of the "fan" diagrams
   thus reducing \eq{lead13} to \eq{sol1}.    This stems from
     the well known fact that the BFKL propagators present 
the leading contribution at high energy in such diagrams.

We find a solution to \eq{sol1} which correctly  describes high energy behavior of the exact solution  and
at the same time satisfies the initial condition at zero rapidity. Its expression is given by \eq{rap4} and has the same
energy structure as the phenomenological "fan" amplitude. The matching between the correct
high energy  behavior and the fulfillment of the initial condition has also another aspect.
It is related to the fact that a strong condition on the source given by \eq{sm7} is not satisfied in general, and
thus dependence on a source breaks the conformal invariance of the solution.
As one can easily see from the form of the solution \eq{rap4} at high energy the dependence on the source
 disappears restoring the conformal invariance.

  As a next step in our discussion we consider a question 
of a transition region in rapidity where a high energy solution
  ansatz transforms into the low energy one. 
This region can be thought of as one where the conformal invariance
  of the theory is restored or where the small 
rapidity expansion can be replaced by the asymptotic expansion. In this
  case some critical conformal scale can be 
introduced through \eq{rap5}-\eq{rap6}. Another way to find the behavior and
  the parametrical form of the critical scale 
is to match between  of low and high energy ansatzs of the solution given by
  \eq{sm8} and \eq{rap4}, respectively. In this 
case the critical scale $z_cr$ may be defined as a scale where the function
  $\bar{g_0}(\nu,z)$ coincides with the function ${g_0}^{-1}$.

It should be mentioned that the proposed ansatz is only an a approximation to the full solutions given by 
series \eq{sol12} and \eq{sm2}
Instead of using ansatz \eq{rap4} and \eq{sm8} one can develop
perturbative calculations and  obtain a chain of equations
similar to ones given by  \eq{sol1} for large rapidities and by  \eq{nsm2}
for of small rapidities.
The calculation of coefficient functions of the expansion \eq{sm2} valid for small rapidities 
corresponds to the calculation of the diagrams in the perturbative expansion. Therefore, the first equation 
( \eq{nsm2} or \eq{nsm4}) from a chain of equations for the coefficient functions is similar to ones obtained in 
Ref. \cite{Hatta} for the simplest "fan" diagrams. The only difference between our result and that of Ref. \cite{Hatta} is the conformal basis and normal coordinate representations correspondingly.
The physical interpretation of the expansion \eq{sol12} is not so clear.  The expansion in the negative powers of 
exponents of rapidity cannot be put into one to one correspondence with diagrams.
The high energy  behavior of the solution to the BK equation in the coordinate 
representation is well known (see \cite{LevT}) and the question of the relation between the conformal 
and coordinate representation at large rapidities will be addressed in our further studies.

 In this paper we considered the expansion of the coordinate 
dependent pomeron fields only in the conformal basis. 
Of course, for the practical applications the coordinate 
representation of the pomeron field is more useful  and
convenient, but the expansion in conformal basis presents a 
more suitable framework for the investigation of the
energy dependence properties of the pomeron field.  
These properties will determine the energy dependence of the full
solution. The task of the inverse transformation of the 
found ansatz into the coordinate basis  we leave
, as we mentioned before, for further publications.

\section*{Acknowledgments}
Authors would like to thank L.Lipatov
for the helpful advises on the subject of the paper. A.P
is grateful to the Santiago de Compostela University and
personally to N.Armesto and C.Pajares for their hospitality
during the stay in Santiago de Compostela. A.P. would like to express his 
deep appreciation to J.Bartels for his hospitality at the University of Hamburg where 
the present work was completed.
    This research was supported in part by the Israel Science Foundation, founded by the Israeli Academy of Science
and Humanities, by a grant from Ministry of Science, Culture and Sport, Israel and the Russian Foundation for Basic
research of the Russian Federation.
This paper was supported by the Ministerio de Educacion y Ciencia of Spain
under project FPA2005-01963, and by Xunta de Galicia
(Conselleria de Educacion).

\vspace{3mm}

\newpage
\section*{Appendix A:}
\renewcommand{\theequation}{A.\arabic{equation}}
\setcounter{equation}{0}

In this Appendix we calculate the integral which appears in  \eq{Addlead3}. To do this we want to rewrite it in terms of some
dimensionless variables and use integral representation of the hypergeometric functions. First we rescale the
variables $z_1$ and $z_2$ as follows

\begin{eqnarray} \label{appA-1}
  \int^{\infty}_{-\infty}  z^{\gamma_1}_1 z^{\gamma_2}_2
(z_0-z_1)^{-\Delta_{\tilde{0}1}}
(z_1-z_2)^{-\Delta_{12}}
(z_2-z_0)^{-\Delta_{2\tilde{0}}} \;d z_1  \;d z_2=(-1)^{-\Delta_{12}-\Delta_{2\tilde{0}}} \bar{z}^{\gamma_1+\gamma_2+2-\Delta_{\tilde{0}1}-\Delta_{12}-\Delta_{2\tilde{0}}}_0 \nonumber
\end{eqnarray}
\begin{eqnarray}
\times \int^{\infty}_{-\infty}
  \left( \frac{z_1}{z_0}\right)^{\gamma_1}
 \left( \frac{z_2}{z_0}\right)^{\gamma_2-\Delta_{12}}
 \left(1-\frac{z_1}{z_0}\right)^{-\Delta_{\tilde{0}1}}
  \left(1-\frac{z_1}{z_0}\frac{z_0}{z_2}\right)^{-\Delta_{12}}
  \left(1-\frac{z_2}{z_0}\right)^{-\Delta_{2\tilde{0}}}
  d \left(\frac{z_1}{z_0} \right)
  d \left(\frac{z_2}{z_0} \right)=    \hspace{1cm}
 \end{eqnarray}
 \begin{eqnarray}
 (-1)^{-\Delta_{12}-\Delta_{2\tilde{0}}} \bar{z}^{\gamma_1+\gamma_2+2-\Delta_{\tilde{0}1}-\Delta_{12}-\Delta_{2\tilde{0}}}_0
 \int^{\infty}_{-\infty}
  w_1^{\gamma_1}
 w_2^{\gamma_2-\Delta_{12}}
 \left(1-w_1\right)^{-\Delta_{\tilde{0}1}}
  \left(1-\frac{w_1}{w_2}\right)^{-\Delta_{12}}
  \left(1-w_2\right)^{-\Delta_{2\tilde{0}}}
  d w_1
  d w_2
  \nonumber
 \end{eqnarray}
where $w_1=z_1/z_0$ and $w_2=z_2/z_0$. Now we perform the integration over $w_1$.
The relevant integral is given by

 \begin{eqnarray} \label{appA-2}
  I_1\equiv \int^{\infty}_{-\infty}
  w_1^{\gamma_1}
 \left(1-w_1\right)^{-\Delta_{\tilde{0}1}}
  \left(1-\frac{w_1}{w_2}\right)^{-\Delta_{12}}
  d w_1
   \end{eqnarray}
 As one can see this integral reminds the integral representation of the Gauss hypergeometric function
 \begin{eqnarray} \label{appA-3}
  \int^{1}_{0}
  t^{b-1} (1-t)^{-b+c-1}(1-tz)^{-a} d t =\frac{\Gamma(b)\Gamma(c-b)}{\Gamma(c)}   \; _2F_1\left(a,b,c,z\right)
  \end{eqnarray}
  but the care about the limits should be taken. We split the limits of the integration in \eq{appA-3} as follows

  \begin{eqnarray} \label{appA-4}
  \int_{-\infty}^{\infty}= \int_{-\infty}^{0}+ \int_{0}^{1}+ \int_{1}^{\infty}
    \end{eqnarray}
    The second term on the rhc in  \eq{appA-4} is just the integral representation of the hypergeometric function given in \eq{appA-3}. The third term  can be brought to the from of  by substitution $w=1/t$ and in this case one
    obtains the following identity
    \begin{eqnarray} \label{appA-5}
  \int_{1}^{\infty}
  t^{b-1} (1-t)^{-b+c-1}(1-tz)^{-a}d t=(-)^{-a-b+c-1}\frac{\Gamma(a-c+1)\Gamma(c-b)}{\Gamma(a-b+1)}
    \; _2F_1\left(a,a-c+1,a-b+1,\frac{1}{z}\right) \hspace{0.3cm}
  \end{eqnarray}
  Thus the integral $(0,\infty)$ is obtained by summing \eq{appA-3} and \eq{appA-5}. The remaining part $(-\infty,0)$ is  obtained from the integral $(0,\infty)$ by substituting $w=1-1/t$ and reads
  \begin{eqnarray} \label{appA-6}
  \int_{-\infty}^{0}
  t^{b-1} (1-t)^{-b+c-1}(1-tz)^{-a}d t=(-)^{-c}\frac{\Gamma(b)\Gamma(a-c)}{\Gamma(a-c+b+1)}
    \; _2F_1\left(a,c-b,a-b+1,\frac{1}{1-z}\right) \hspace{0.3cm}
  \end{eqnarray}
    Summing the expressions of \eq{appA-3}, \eq{appA-5} and \eq{appA-6} we can write the full expression to be used
    for $w_1$ integration as follows.

     \begin{eqnarray} \label{appA-7}
    \int_{-\infty}^{\infty}
  t^{b-1} (1-t)^{-b+c-1}(1-tz)^{-a}d t= (-)^{-c}\frac{\Gamma(b)\Gamma(a-c)}{\Gamma(a-c+b+1)}
    \; _2F_1\left(a,c-b,a-b+1,\frac{1}{1-z}\right) +
    \end{eqnarray}
     \begin{eqnarray}
    + (-)^{-a-b+c-1}\frac{\Gamma(a-c+1)\Gamma(c-b)}{\Gamma(a-b+1)}
    \; _2F_1\left(a,a-c+1,a-b+1,\frac{1}{z}\right)
     +
      \frac{\Gamma(b)\Gamma(c-b)}{\Gamma(c)}   \; _2F_1\left(a,b,c,z\right)\nonumber
     \end{eqnarray}

     Comparing \eq{appA-2} and \eq{appA-7} we readily identify the parameters of  \eq{appA-7} as
     \begin{eqnarray} \label{appA-8}
      a=\Delta_{12} \; , \hspace{1cm} b=1+\gamma_1 \; , \hspace{1cm}
      c=\gamma_1-\Delta_{\tilde{0}1}+2 \;,\hspace{1cm}
      z=\frac{1}{w_2} \hspace{1cm}
     \end{eqnarray}
Because of the inverse dependence of $w_2$ on $z$ in \eq{appA-8} it
is more convenient for the further integration over $w_2$ to rewrite
\eq{appA-7} in terms of the hypergeometric functions of the same
argument $\frac{1}{z}$. This can be done using useful identities for
the hypergeometric functions as follows ( see \textbf{15.3.4} and
\textbf{15.3.7} in \cite{STEGUN})

\begin{eqnarray} \label{appA-9}
      \; _2F_1(a,b,c;\frac{1}{1-z})=(-1)^a(1-z)^a z^{-a}
      \; _2F_1(a,c-b,c;\frac{1}{z})
     \end{eqnarray}
     and
     \begin{eqnarray} \label{appA-10}
      \; _2F_1(a,b,c;z)&=&(-1)^a z^{-a}\frac{\Gamma(c)\Gamma(b-a)}{\Gamma(b)\Gamma(c-a)}
      \; _2F_1(a,1-c+a,1-b+a;\frac{1}{z})\\ &&+
      (-1)^b z^{-b}\frac{\Gamma(c)\Gamma(a-b)}{\Gamma(a)\Gamma(c-b)}
      \; _2F_1(b,1-c+b,1-a+b;\frac{1}{z}) \nonumber
     \end{eqnarray}
With the help of \eq{appA-9} and \eq{appA-10} the integral in
\eq{appA-7} reads

\begin{eqnarray} \label{appA-11}
    \int_{-\infty}^{\infty}
  t^{b-1} (1-t)^{-b+c-1}(1-tz)^{-a}d t= C_1 \cdot
  \; _2F_1\left(b,b-c+1,-a+b+1,\frac{1}{z}\right)
  + C_2 \cdot\; _2F_1\left(a,a-c+1,a-b+1,\frac{1}{z}\right)
   \nonumber
    \end{eqnarray}
with the functions $C_1$ and $C_2$ given by
\begin{eqnarray} \label{appA-12}
    C_1=(-1)^b \left(\frac{1}{z}\right)^b \frac{\Gamma(b)\Gamma(c-b)}{\Gamma(c)}  \nonumber
    \end{eqnarray}
and
\begin{eqnarray} \label{appA-13}
C_2= (-1)^{-a-b+c-1}\frac{\Gamma(a-c+1)\Gamma(c-b)}{\Gamma(a-b+1)} +
  (-1)^a \left(\frac{1}{z}\right)^{a} \frac{\Gamma(c-b)\Gamma(b-a)}{\Gamma(c-a)}
      +
  (-1)^{c}\left(1-\frac{1}{z}\right)^a
  \frac{\Gamma(b)\Gamma(a-c)}{\Gamma(a+b-c+1)}
   \nonumber
    \end{eqnarray}
Thus with the help of \eq{appA-8} we identify the required integral \eq{appA-2} as
\begin{eqnarray}\label{appA-111}
I_1=C_1 \cdot
  \; _2F_1\left(1+\gamma_1,\Delta_{\tilde{0}1},2+\Delta_{12}+\gamma_1,w_2\right)
  + C_2 \cdot\; _2F_1\left(\Delta_{12},\Delta_{12}-\gamma_1+\Delta_{\tilde{0},1}-1,\Delta_{12}-\gamma_1,w_2\right)
  \hspace{1cm}
\end{eqnarray}
with the functions $C_1$ and $C_2$ given by
\begin{eqnarray} \label{appA-112}
    C_1=(-1)^{1+\gamma_1} w_2^{1+\gamma_1} \frac{\Gamma(1+\gamma_1)
    \Gamma(1-\Delta_{\tilde{0}1})}{\Gamma(\gamma_1-\Delta_{\tilde{0}1}+2)}  \nonumber
    \end{eqnarray}
     and
\begin{eqnarray} \label{appA-113}
C_2 &=& (-1)^{\Delta_{\tilde{0}1}+\Delta_{12}}\frac{\Gamma(\Delta_{\tilde{0}1}+\Delta_{12}-\gamma_1-1)
\Gamma(1-\Delta_{\tilde{0}1})}{\Gamma(\Delta_{12}-\gamma_1)} +
  (-1)^{\Delta_{12}} w_2^{\Delta_{12}} \frac{\Gamma(1-\Delta_{\tilde{0}1})
  \Gamma(1+\gamma_1-\Delta_{12})}{\Gamma(2+\gamma_2-\Delta_{\tilde{0}1}-\Delta_{12})} \nonumber
  \\
     &&+
  (-1)^{\gamma_1-\Delta_{\tilde{0}1}+2}\left(1-w_2\right)^{\Delta_{12}}
  \frac{\Gamma(1+\gamma_1)\Gamma(\Delta_{\tilde{0}1}+\Delta_{12}-\gamma_1-2)}{\Gamma(\Delta_{\tilde{0}1}+\Delta_{12})}
    \end{eqnarray}

The next step is to perform integration over variable $w_2$.
   From \eq{appA-1} with the definition of
 \eq{appA-2} we see that the  integral over $w_2$ reads

\begin{eqnarray} \label{appA-14}
  \int^{\infty}_{-\infty} I_1
  w_2^{\gamma_2-\Delta_{12}}
 \left(1-w_2\right)^{-\Delta_{2\tilde{0}}}
  d w_2
   \end{eqnarray}
It is clear from \eq{appA-14} and the result of the integration over
$w_1$ that the relevant integral is
\begin{eqnarray} \label{appA-15}
\int_{-\infty}^{\infty} w^{\alpha} (1-w)^{\beta} \;
_2F_1\left(a,b,c,w\right)dw
\end{eqnarray}
As in the case of the integration over $w_1$ we want to use the
identity (see \textbf{7.152.5} in \cite{Ryzhik})
\begin{eqnarray} \label{appA-16}
\int_{0}^{1} w^{\alpha} (1-w)^{\beta} \;
_2F_1\left(a,b,c,w\right)dw=
\frac{\Gamma(\alpha+1)\Gamma(\beta+1)}{\Gamma(\alpha+\beta+2)}
\;_3F_2\left(a,b,\alpha+1;c,\alpha+\beta+2;1\right)
\end{eqnarray}
and thus split the integration in \eq{appA-15} as in \eq{appA-4}.
The integral $(1,\infty)$ is obtained from \eq{appA-16} by
substitution $w \rightarrow 1/w$ and with the help of the identity
\eq{appA-10}. The result reads
\begin{eqnarray} \label{appA-17}
&&\int_{1}^{\infty} w^{\alpha} (1-w)^{\beta} \;
_2F_1\left(a,b,c,w\right)dw=\\&&(-1)^{a+\beta}
\frac{\Gamma(a-\alpha-\beta-1)\Gamma(\beta+1)}{\Gamma(a-\alpha)}
\frac{\Gamma(c)\Gamma(b-a)}{\Gamma(b)\Gamma(c-a)}
\;_3F_2\left(a,a-c+1,a-\alpha-\beta-1;a-b+1,a-\alpha;1\right) +
\nonumber\\&& (-1)^{b+\beta}
\frac{\Gamma(b-\alpha-\beta-1)\Gamma(\beta+1)}{\Gamma(b-\alpha)}
\frac{\Gamma(c)\Gamma(a-b)}{\Gamma(a)\Gamma(c-b)}
\;_3F_2\left(b,b-c+1,b-\alpha-\beta-1;-a+b+1,b-\alpha;1\right)
\nonumber
\end{eqnarray}
The sum of \eq{appA-16} and  \eq{appA-17} gives the contribution
from the integration $(0,\infty)$. As in \eq{appA-6} the missing
part $(-\infty,0)$ is obtained from integral $(0,1)$ substituting $w
\rightarrow 1/(1-w)$ and using identities \eq{appA-9} and
\eq{appA-10}
\begin{eqnarray} \label{appA-18}
&&\int_{-\infty}^{0} w^{\alpha} (1-w)^{\beta} \;
_2F_1\left(a,b,c,w\right)dw=\\&&(-1)^{\alpha}
\frac{\Gamma(a-\alpha-\beta-1)\Gamma(\alpha+1)}{\Gamma(a-\beta)}
\frac{\Gamma(c)\Gamma(b-c)}{\Gamma(b)\Gamma(c-a)}
\;_3F_2\left(a,b-c,a-\alpha-\beta-1;-a-b+1,a-\beta;1\right) +
\nonumber\\&& (-1)^{\alpha}
\frac{\Gamma(b-\alpha-\beta-1)\Gamma(\alpha+1)}{\Gamma(b-\beta)}
\frac{\Gamma(c)\Gamma(a-b)}{\Gamma(a)\Gamma(c-b)}
\;_3F_2\left(-b+c,a,b-\alpha-\beta-1;c,b-\beta;1\right) \nonumber
\end{eqnarray}
Finally summing the contributions of \eq{appA-16}, \eq{appA-17} and
\eq{appA-18} and plugging it into \eq{appA-15} we obtain the
contribution to \eq{appA-1} coming from the integration over the
holomorphic variables $z_1$ and $z_2$.
The corresponding expression reads

\begin{eqnarray}
-(-1)^{\Delta_{\tilde{0}1}-\Delta_{12}}
 B[1+\gamma_2-\Delta_{12},1-\Delta_{2\tilde{0}}] B[1+\Delta_{12},\Delta_{\tilde{0}1}-\Delta_{12}]\\
\times\;_3F_2[\{\gamma_1,-\Delta_{\tilde{0}1},1+\gamma_2-\Delta_{12}\},
\{-\Delta_{12},2+\gamma_2-\Delta_{12}-\Delta_{2\tilde{0}}\},1]  \nonumber
\end{eqnarray}
\begin{eqnarray} \nonumber
-(-1)^{\Delta_{\tilde{0}1}-\Delta_{12}}
 B[1+\gamma_2-\Delta_{12},1-\Delta_{2\tilde{0}}] B(1+\Delta_{12},\Delta_{\tilde{0}1}-\Delta_{12}]
 \\
\times\;_3F_2[\{\gamma_1,-\Delta_{\tilde{0}1},1+\gamma_2-\Delta_{12}\},
\{-\Delta_{12},2+\gamma_2-\Delta_{12}-\Delta_{2\tilde{0}}\},1]
\nonumber \end{eqnarray}\begin{eqnarray} \nonumber
-(-1)^{-\Delta_{\tilde{0}1}+\Delta_{12}}
B(1+\gamma_2-\Delta_{12},1-\Delta_{2\tilde{0}}]
B(1+\Delta_{12},\Delta_{\tilde{0}1}-\Delta_{12}]
\\
 \times\;_3F_2(\{\gamma_1,-\Delta_{\tilde{0}1},1+\gamma_2-\Delta_{12}\},\{-\Delta_{12},2+\gamma_2-\Delta_{12}-\Delta_{2\tilde{0}}\},1]
\nonumber \end{eqnarray}\begin{eqnarray} \nonumber
-(-1)^{\gamma_2-\Delta_{\tilde{0}1}}
B(1+\gamma_2-\Delta_{12},-1-\gamma_2+\Delta_{12}+\Delta_{2\tilde{0}}]
B(1+\Delta_{12},\Delta_{\tilde{0}1}-\Delta_{12}]
  \\
\times\;_3F_2(\{\gamma_1,-\Delta_{\tilde{0}1},1+\gamma_2-\Delta_{12}\},\{-\Delta_{12},2+\gamma_2-\Delta_{12}-\Delta_{2\tilde{0}}\},1]
\nonumber \end{eqnarray}\begin{eqnarray} \nonumber
-(-1)^{\gamma_2+\Delta_{\tilde{0}1}-2 \Delta_{12}}
B(1-\gamma_1+\gamma_2-\Delta_{12},-1+\gamma_1-\gamma_2+\Delta_{12}+\Delta_{2\tilde{0}}]
B(1+\gamma_1+\Delta_{12},\Delta_{\tilde{0}1}-\Delta_{12}] \\
\times\;_3F_2(\{\gamma_1,1+\gamma_1+\Delta_{12},-1+\gamma_1-\gamma_2+\Delta_{12}+\Delta_{2\tilde{0}}\},\{1+\gamma_1+\Delta_{\tilde{0}1},\gamma_1-\gamma_2+\Delta_{12}\},1]
\nonumber \end{eqnarray}\begin{eqnarray} \nonumber
-\frac{(-1)^{\gamma_1-\Delta_{\tilde{0}1}-2
(\gamma_2-\Delta_{12})-\Delta_{12}-\Delta_{2\tilde{0}}}
B(-\gamma_1-\Delta_{\tilde{0}1},1+\gamma_1+\Delta_{12}]
B(1+\Delta_{12},-\Delta_{12}]
B(-1+\gamma_1-\gamma_2+\Delta_{12}+\Delta_{2\tilde{0}},1-\Delta_{2\tilde{0}}]
}{B(1-\gamma_1,\gamma_1]} \\ \times\;_3F_2(\{\gamma_1,1+\gamma_1+\Delta_{12},-1+\gamma_1-\gamma_2+\Delta_{12}+\Delta_{2\tilde{0}}\},\{1+\gamma_1+\Delta_{\tilde{0}1},\gamma_1-\gamma_2+\Delta_{12}\},1]
\nonumber \end{eqnarray}\begin{eqnarray} \nonumber
-\frac{(-1)^{\gamma_1-\Delta_{\tilde{0}1}-2
(\gamma_2-\Delta_{12})+\Delta_{12}-\Delta_{2\tilde{0}}}
B(-\gamma_1-\Delta_{\tilde{0}1},\Delta_{\tilde{0}1}-\Delta_{12}]
B(1+\Delta_{12},-\Delta_{12}]
B(-1+\gamma_1-\gamma_2+\Delta_{12}+\Delta_{2\tilde{0}},1-\Delta_{2\tilde{0}}]
 }{B(1+\Delta_{\tilde{0}1},-\Delta_{\tilde{0}1}]}
  \\
  \times\;_3F_2(\{\gamma_1,1+\gamma_1+\Delta_{12},-1+\gamma_1-\gamma_2+\Delta_{12}+\Delta_{2\tilde{0}}\},\{1+\gamma_1+\Delta_{\tilde{0}1},\gamma_1-\gamma_2+\Delta_{12}\},1]
\nonumber \end{eqnarray}\begin{eqnarray} \nonumber
-(-1)^{-\gamma_1+\Delta_{\tilde{0}1}-\Delta_{12}-\Delta_{2\tilde{0}}}
B(1+\gamma_1+\Delta_{12},\Delta_{\tilde{0}1}-\Delta_{12}]
B(-1+\gamma_1-\gamma_2+\Delta_{12}+\Delta_{2\tilde{0}},1-\Delta_{2\tilde{0}}] \\
 \times\;_3F_2(\{\gamma_1,1+\gamma_1+\Delta_{12},-1+\gamma_1-\gamma_2+\Delta_{12}+\Delta_{2\tilde{0}}\},\{1+\gamma_1+\Delta_{\tilde{0}1},\gamma_1-\gamma_2+\Delta_{12}\},1]
\nonumber \end{eqnarray}\begin{eqnarray} \nonumber
-(-1)^{\gamma_2+\Delta_{\tilde{0}1}-2 \Delta_{12}}
B(\Delta_{\tilde{0}1}-\Delta_{12},2+\gamma_2]
B(-1+\gamma_1-\gamma_2+\Delta_{12},1+\gamma_2-\Delta_{12}] \\
 \times\;_3F_2(\{2+\gamma_2,1+\gamma_2-\Delta_{12},\Delta_{2\tilde{0}}\},\{2-\gamma_1+\gamma_2-\Delta_{12},2+\gamma_2+\Delta_{\tilde{0}1}-\Delta_{12}\},1]
\nonumber \end{eqnarray}\begin{eqnarray} \nonumber
+\frac{(-1)^{-\Delta_{\tilde{0}1}}
B(2+\gamma_2,1-\Delta_{2\tilde{0}}] B(1+\Delta_{12},-\Delta_{12}]
}{(1+\gamma_1+\Delta_{12})
B(\gamma_1,2+\Delta_{12}]}
\\
\times\;_3F_2(\{2+\gamma_2,1+\gamma_1+\Delta_{12},1-\Delta_{\tilde{0}1}+\Delta_{12}\},\{2+\Delta_{12},3+\gamma_2-\Delta_{2\tilde{0}}\},1]
\nonumber \end{eqnarray}\begin{eqnarray} \nonumber
-\frac{(-1)^{\gamma_2-\Delta_{\tilde{0}1}}
B(2+\gamma_2,-2-\gamma_2+\Delta_{2\tilde{0}}]
B(1+\Delta_{12},-\Delta_{12}]
}{(1+\gamma_1+\Delta_{12})
B(\gamma_1,2+\Delta_{12}]}
\\
\times\;_3F_2(\{2+\gamma_2,1+\gamma_1+\Delta_{12},1-\Delta_{\tilde{0}1}+\Delta_{12}\},\{2+\Delta_{12},3+\gamma_2-\Delta_{2\tilde{0}}\},1]
\nonumber \end{eqnarray}\begin{eqnarray} \nonumber
+(-1)^{\Delta_{\tilde{0}1}} B(2+\gamma_2,1-\Delta_{2\tilde{0}}]
B(1+\gamma_1+\Delta_{12},-1-\Delta_{12}] \\
\times\;_3F_2(\{2+\gamma_2,1+\gamma_1+\Delta_{12},1-\Delta_{\tilde{0}1}+\Delta_{12}\},\{2+\Delta_{12},3+\gamma_2-\Delta_{2\tilde{0}}\},1]
\nonumber \end{eqnarray}\begin{eqnarray} \nonumber
-\frac{(-1)^{-2 \Delta_{\tilde{0}1}-2
(\gamma_2-\Delta_{12})-\Delta_{12}-\Delta_{2\tilde{0}}}
B(1+\Delta_{12},-\Delta_{12}]
B(-1-\gamma_2-\Delta_{\tilde{0}1}+\Delta_{12}+\Delta_{2\tilde{0}},1-\Delta_{2\tilde{0}}]
}{(\gamma_1+\Delta_{\tilde{0}1})
B(\gamma_1,1+\Delta_{\tilde{0}1}]}
\\
\times\;_3F_2(\{-\Delta_{\tilde{0}1},1-\Delta_{\tilde{0}1}+\Delta_{12},-1-\gamma_2-\Delta_{\tilde{0}1}+\Delta_{12}+\Delta_{2\tilde{0}}\},\{1-\gamma_1-\Delta_{\tilde{0}1},-\gamma_2-\Delta_{\tilde{0}1}+\Delta_{12}\},1]
\nonumber \end{eqnarray}\begin{eqnarray} \nonumber
+\frac{(-1)^{-2 \Delta_{\tilde{0}1}-2
(\gamma_2-\Delta_{12})+\Delta_{12}-\Delta_{2\tilde{0}}}
B(1+\Delta_{12},-\Delta_{12}]
B(-1-\gamma_2-\Delta_{\tilde{0}1}+\Delta_{12}+\Delta_{2\tilde{0}},1-\Delta_{2\tilde{0}}]
}{(\gamma_1+\Delta_{\tilde{0}1})
B(\gamma_1,1+\Delta_{\tilde{0}1}]}
\\
\times\;_3F_2(\{-\Delta_{\tilde{0}1},1-\Delta_{\tilde{0}1}+\Delta_{12},-1-\gamma_2-\Delta_{\tilde{0}1}+\Delta_{12}+\Delta_{2\tilde{0}}\},\{1-\gamma_1-\Delta_{\tilde{0}1},-\gamma_2-\Delta_{\tilde{0}1}+\Delta_{12}\},1]
\nonumber \end{eqnarray}\begin{eqnarray}
-(-1)^{\gamma_2-\Delta_{12}}
B(\Delta_{\tilde{0}1}-\Delta_{12},-1-\gamma_2-\Delta_{\tilde{0}1}+\Delta_{12}+\Delta_{2\tilde{0}}]
B(-1+\gamma_1-\gamma_2+\Delta_{12}+\Delta_{2\tilde{0}},1+\gamma_2-\Delta_{12}-\Delta_{2\tilde{0}}]\nonumber
\\
\times\;_3F_2(\{\Delta_{2\tilde{0}},-1+\gamma_1-\gamma_2+\Delta_{12}+\Delta_{2\tilde{0}},-1-\gamma_2-\Delta_{\tilde{0}1}+\Delta_{12}+\Delta_{2\tilde{0}}\},\{-1-\gamma_2+\Delta_{2\tilde{0}},-\gamma_2+\Delta_{12}+\Delta_{2\tilde{0}}\},1]
\nonumber \end{eqnarray}\begin{eqnarray} \nonumber
-\frac{(-1)^{\gamma_2-\Delta_{\tilde{0}1}-\Delta_{12}}
B(1+\Delta_{12},-\Delta_{12}]
B(-1+\gamma_1-\gamma_2+\Delta_{12}+\Delta_{2\tilde{0}},1+\gamma_2-\Delta_{12}-\Delta_{2\tilde{0}}]
}{(-1-\gamma_2-\Delta_{\tilde{0}1}+\Delta_{12}+\Delta_{2\tilde{0}})
B(1-\Delta_{\tilde{0}1}+\Delta_{12},-1-\gamma_2+\Delta_{2\tilde{0}}]}
\\
\times\;_3F_2(\{\Delta_{2\tilde{0}},-1+\gamma_1-\gamma_2+\Delta_{12}+\Delta_{2\tilde{0}},-1-\gamma_2-\Delta_{\tilde{0}1}+\Delta_{12}+\Delta_{2\tilde{0}}\},\{-1-\gamma_2+\Delta_{2\tilde{0}},-\gamma_2+\Delta_{12}+\Delta_{2\tilde{0}}\},1]
\nonumber \end{eqnarray}\begin{eqnarray} \nonumber
-\frac{(-1)^{\gamma_2-\Delta_{\tilde{0}1}}
B(1+\Delta_{12},-\Delta_{12}]
B(-1-\gamma_2-\Delta_{\tilde{0}1}+\Delta_{12}+\Delta_{2\tilde{0}},2+\gamma_2-\Delta_{2\tilde{0}}]
}{(-1+\gamma_1-\gamma_2+\Delta_{12}+\Delta_{2\tilde{0}})
B(\gamma_1,-\gamma_2+\Delta_{12}+\Delta_{2\tilde{0}}]}
\\
\times\;_3F_2(\{\Delta_{2\tilde{0}},-1+\gamma_1-\gamma_2+\Delta_{12}+\Delta_{2\tilde{0}},-1-\gamma_2-\Delta_{\tilde{0}1}+\Delta_{12}+\Delta_{2\tilde{0}}\},\{-1-\gamma_2+\Delta_{2\tilde{0}},-\gamma_2+\Delta_{12}+\Delta_{2\tilde{0}}\},1]
\nonumber \end{eqnarray}

The integration over antiholomorphic
variables $\bar{z}_1$ and $\bar{z}_2$ now can be easily performed using this last result.


\newpage
\begin{thebibliography}{100}



\bibitem{vert1}
J.~Bartels,
Z.\ Phys.\ C {\bf 60} (1993) 471.

\bibitem{vert2}
J.~Bartels and M.~W\"{u}sthoff,
Z.\ Phys.\ C {\bf 66} (1995) 157.


\bibitem{brav}
  M.~Braun and G.~P.~Vacca,
  Eur.\ Phys.\ J.\  C {\bf 4}, (1998) 85.


\bibitem{Schwimmer}
  A.~Schwimmer,
  Nucl.\ Phys.\  B {\bf 94}, (1975) 445.


\bibitem{GLR}
  L.~V.~Gribov, E.~M.~Levin and M.~G.~Ryskin,
  Phys.\ Rept.\  {\bf 100}, (1983) 1.



\bibitem{bfkl1}
  L.~N.~Lipatov,
  Sov.\ J.\ Nucl.\ Phys.\  {\bf 23} (1976) 338
  [Yad.\ Fiz.\  {\bf 23} (1976) 642];

  E.~A.~Kuraev, L.~N.~Lipatov and V.~S.~Fadin,
  Sov.\ Phys.\ JETP {\bf 45} (1977) 199
  [Zh.\ Eksp.\ Teor.\ Fiz.\  {\bf 72} (1977) 377];

  I.~I.~Balitsky and L.~N.~Lipatov,
  Sov.\ J.\ Nucl.\ Phys.\  {\bf 28} (1978) 822
  [Yad.\ Fiz.\  {\bf 28} (1978) 1597].


\bibitem{bfkl2}
L.~N.~Lipatov, {Phys.\ Rept.\ } {\bf  286} (1997) 131.


\bibitem{bfkl3}
  V.~S.~Fadin and L.~N.~Lipatov,
  Phys.\ Lett.\ B {\bf 429} (1998) 127;

M.~Ciafaloni and G.~Camici,
  Phys.\ Lett.\ B {\bf 430} (1998) 349;

  V.~S.~Fadin and R.~Fiore,
  Phys.\ Lett.\ B {\bf 610} (2005) 61
  [Erratum-ibid.\ B {\bf 621} (2005) 61];

  V.~S.~Fadin and R.~Fiore,
  Phys.\ Rev.\ D {\bf 72} (2005) 014018.



\bibitem{bal} I.I.Balitsky, Nucl. Phys. {\bf B463} (1996) 99.
%
\bibitem{kov} Yu.V.Kovchegov, Phys. Rev. {\bf D60} (1999) 034008;
{\bf D61} (2000) 074018.

\bibitem{muel}
A.~H.~Mueller,
Nucl.\ Phys.\ B {\bf 415} (1994) 373.


\bibitem{bksols}
  E.~Levin and K.~Tuchin,
  Nucl.\ Phys.\ B {\bf 573} (2000) 833;

M.~Braun,
Eur.\ Phys.\ J.\ C {\bf 16} (2000) 337;

  H.~Weigert,
  Nucl.\ Phys.\ A {\bf 703} (2002) 823;

  N.~Armesto and M.~A.~Braun,
  Eur.\ Phys.\ J.\ C {\bf 20} (2001) 517;

  K.~Golec-Biernat, L.~Motyka and A.~M.~Sta\'{s}to,
  Phys.\ Rev.\ D {\bf 65} (2002) 074037;

G.~Chachamis, M.~Lublinsky and A.~Sabio Vera,
Nucl.\ Phys.\ A {\bf 748} (2005) 649.


\bibitem{bdepkov}
  K.~Golec-Biernat and A.~M.~Sta\'{s}to,
  Nucl.\ Phys.\ B {\bf 668} (2003) 345.



\bibitem{jimsol}
  K.~Rummukainen and H.~Weigert,
  Nucl.\ Phys.\ A {\bf 739} (2004) 183.




\bibitem{bksemi1}
  E.~Levin and K.~Tuchin,
  Nucl.\ Phys.\ A {\bf 691} (2001) 779;

  J.~Kwieci\'{n}ski and A.~M.~Sta\'{s}to,
  Phys.\ Rev.\ D {\bf 66} (2002) 014013;

  E.~Iancu, K.~Itakura and L.~McLerran,
  Nucl.\ Phys.\ A {\bf 708} (2002) 327;

  S.~Bondarenko, M.~Kozlov and E.~Levin,
  Nucl.\ Phys.\  A {\bf 727} (2003) 139 .



\bibitem{bksemi2}
  A.~H.~Mueller and D.~N.~Triantafyllopoulos,
  Nucl.\ Phys.\ B {\bf 640} (2002) 331;

  D.~N.~Triantafyllopoulos,
  Nucl.\ Phys.\ B {\bf 648} (2003) 293;


  L.~Motyka, Phys.\ Lett.\ B ,
  arXiv:hep-ph/0509270.


\bibitem{barlev}
  J.~Bartels and E.~Levin,
  Nucl.\ Phys.\  B {\bf 387} (1992) 617.


\bibitem{scaling}
  A.~M.~Sta\'{s}to, K.~Golec-Biernat and J.~Kwieci\'{n}ski,
  Phys.\ Rev.\ Lett.\  {\bf 86} (2001) 596.


\bibitem{traveling}
  S.~Munier and R.~Peschanski,
  Phys.\ Rev.\ Lett.\  {\bf 91} (2003) 232001;

  S.~Munier and R.~Peschanski,
  Phys.\ Rev.\ D {\bf 69}, 034008 (2004);

  S.~Munier and R.~Peschanski,
  Phys.\ Rev.\ D {\bf 70}, 077503 (2004).




  \bibitem{bon1}
  S.~Bondarenko, E.~Gotsman, E.~Levin and U.~Maor,
  Nucl.\ Phys.\  A {\bf 683} (2001) 649.




\bibitem{braun1}
  M.~A.~Braun,
  Phys.\ Lett.\ B {\bf 483} (2000) 115.

\bibitem{braun2}
  M.~A.~Braun,
  Eur.\ Phys.\ J.\ C {\bf 33} (2004) 113.

\bibitem{braun3}
M.~A.~Braun,
arXiv:hep-ph/0504002.

\bibitem{braun4}
M.~A.~Braun,
Phys.\ Lett.\ B {\bf 632} (2006) 297.


\bibitem{bom}
  S.~Bondarenko and L.~Motyka,
  Phys.\ Rev.\  D {\bf 75}, (2007) 114015.


\bibitem{bon2}
  S.~Bondarenko,
  Nucl.\ Phys.\  A {\bf 792} (2007) 264.




\bibitem{lip1}
L.N.Lipatov, in "Perturbative QCD", ed. A.H.Mueller,
World. Sci. Singapore (1989).

\bibitem{Ham1}
L.N.Lipatov,
Sov.\ Phys.\ JETP {\bf 63} (1986) 904,
Nucl. \ Phys. \ B {\bf 715} (1991) 641,
Phys. \ Rept.\  {\bf 286} (1997) 131.

\bibitem{Ham2}
J.~Bartels, L.~N.~Lipatov and G.~P.~Vacca,
Nucl.\ Phys.\ B {\bf 706} (2005) 391.


\bibitem{pesc}
  R.~Peschanski,
  Phys.\ Lett.\  B {\bf 409} (1997) 491.

\bibitem{korm}
  G.~P.~Korchemsky,
  Nucl.\ Phys.\  B {\bf 550} (1999) 397.


\bibitem{Hatta}
  Y.~Hatta and A.~H.~Mueller,
  Nucl.\ Phys.\  A {\bf 789} (2007) 285.


\bibitem{LevT}
  E.~Levin and K.~Tuchin,
  Nucl.\ Phys.\  B {\bf 573} (2000) 833;

  E.~Levin and K.~Tuchin,
  Nucl.\ Phys.\  A {\bf 691} (2001) 779.

\bibitem{STEGUN}
Abramowitz, M., and  Stegun, I. S. 1972, Handbook of Mathematical Functions
  (New York: Dover)

\bibitem{Ryzhik}
I. S. Gradstein and I. M. Ryzhik, in Tables of Integrals, Series and Products, edited by A. Jeffrey (Academic, New York, 1980)





\end{thebibliography}
\end{document}